\documentclass[conference,12pt]{IEEEtran}
\usepackage{cite}
\usepackage[T1]{fontenc}
\usepackage[utf8]{inputenc}
\usepackage{url}
\usepackage{hyperref}
\usepackage{graphicx}
\usepackage[table]{xcolor}
\usepackage{array}
\usepackage[most]{tcolorbox}
\usepackage{cuted}
\usepackage{capt-of}
\usepackage{flushend}
\begin{document}

\title{Thinking Out Loud: Real-Time Deception Monitoring in Asymmetric LLM Negotiations}

\author{
\IEEEauthorblockN{Nolan Coffey}
\IEEEauthorblockA{University of Tennessee\\
Knoxville, Tennessee\\
ncoffey3@vols.utk.edu}
\and
\IEEEauthorblockN{Faithful Odoi}
\IEEEauthorblockA{University of Tennessee\\
Knoxville, Tennessee\\
cvy221@vols.utk.edu}
\and
\IEEEauthorblockN{Makenzie Johnson}
\IEEEauthorblockA{University of Tennessee\\
Knoxville, Tennessee\\
mjohn326@vols.utk.edu}
\and
\IEEEauthorblockN{Nasir U. Eisty}
\IEEEauthorblockA{University of Tennessee\\
Knoxville, Tennessee\\
neisty@utk.edu}
}
\maketitle

\begin{abstract}
As LLM-based agents are increasingly deployed to negotiate, delegate, or transact on a user's behalf, software pipelines need runtime mechanisms to verify that an agent's stated intentions match its actual behavior. We study whether a lightweight, real-time chain-of-thought (CoT) monitor can detect strategic deception during asymmetric negotiations, using a used-car sales scenario where a seller agent has private knowledge of an undisclosed defect and a buyer agent has only public market data. The monitor, implemented as a third agent, audits the seller's internal reasoning against its messages and alerts the buyer whenever concealment is detected, across multiple buyer-seller model pairings. Our experiments show that this monitor increases the buyer's walk-away rate, but reveal a persistent intelligence gap: lower-capability buyers often cannot translate an alert into an equitable counter-offer and still accept exploitative deals after being warned. Sellers also change their behavior when told they are monitored, though concealment is not eliminated. These results highlight both the promise and limits of lightweight real-time oversight, offering practical guidance for engineers building and validating monitoring infrastructure for agentic systems with conflicting stakeholder incentives.
\end{abstract}

\section{Introduction}

Software systems are increasingly built around autonomous agents powered by LLMs: coding assistants that open pull requests, DevOps agents that provision infrastructure, and negotiation or procurement agents that interact with external services and with each other on a user's behalf. As these agents are composed into larger pipelines, the correctness of the overall system depends not only on each agent producing the right output, but on each agent behaving honestly toward the other agents and stakeholders it interacts with. This raises a software engineering question closely related to, but distinct from, the AI safety question of alignment: how should engineers design, implement, and validate monitoring components that detect when an agent's communicated behavior diverges from its internal reasoning?

LLMs continue to grow in capability and are increasingly deployed as autonomous agents trusted to act on behalf of human stakeholders. Prior work has shown that as model capability increases, so does the willingness of LLM agents to engage in misaligned behavior, including simulated blackmail and espionage \cite{lynch2025agentic}, and that LLMs will act deceptively to achieve strategic goals \cite{su2025ailiedar}. As models become more capable, our ability to keep them aligned with the Helpful, Honest, and Harmless (HHH) principles \cite{askell2021general} must keep pace. One proposed approach is weak-to-strong generalization, in which a weaker, aligned model supervises a stronger one \cite{burns2024weak}. However, prior work suggests that stronger models can exploit weak supervision when their misaligned behavior is difficult for the supervisor to detect \cite{yang2025superficial}. This motivates real-time monitoring systems that can assist users, or weaker agents, during an interaction itself, rather than only evaluating behavior after the fact.

Monitoring systems have been proposed for both model internals and chain-of-thought (CoT) reasoning traces \cite{shah2025approach}. CoT monitoring is especially relevant in agentic settings, since an agent's reasoning can reveal intentions or contradictions not visible from its final response alone \cite{baker2025cot}. In this work, we examine whether a lightweight CoT-based deception monitor can identify, and help a counterpart respond to, concealment during an asymmetric negotiation.

We study this question through a realistic used-car negotiation in which a seller agent holds private adverse information about the vehicle's transmission condition, while the buyer agent has access only to public market information. This asymmetry gives the seller both the opportunity and the incentive to conceal material information in pursuit of a better outcome. We evaluate the negotiation across multiple buyer-seller model pairings, with and without a real-time deception monitor that alerts the buyer whenever it detects concealment.

This work contributes a focused empirical study of deception monitoring in asymmetric LLM negotiation. Rather than asking only whether LLMs can deceive, we examine how often concealment occurs in a realistic negotiation, how model pairings shape negotiation outcomes, whether real-time monitoring changes those outcomes, and whether sellers behave differently when they know they are being monitored. We hope this provides insight into the safe deployment of agentic AI systems in high-stakes interactions where agents represent stakeholders with opposing goals. From a software engineering perspective, our negotiation environment, monitoring component, and validation methodology also serve as a template for testing and auditing oversight mechanisms before they are integrated into larger multi-agent pipelines.

Specifically, this work aims to answer the following research questions:

\noindent\textbf{RQ1:} How often and in what forms do seller agents engage in deception when given private adverse information in a negotiation?\\
\noindent\textbf{RQ2:} How does capability asymmetry between buyer and seller models affect negotiation outcomes, deceptive behavior, and the buyer's ability to resist exploitation?\\
\noindent\textbf{RQ3:} Can deception detection methods be deployed in real time to help counter misaligned behavior by a seller model?\\
\noindent\textbf{RQ3a:} Does an agent behave differently if it knows it is being monitored?\\

\section{Related Work}

As the development of LLMs continues to accelerate, the study of multi-agent interaction, deception, and oversight has become an increasingly important area of research. Our work differs from prior studies by investigating how LLMs interact in asymmetric negotiation contexts, while also examining whether deception detection methods can be deployed in real time to counter misaligned behavior.

The Deception Detection subsection covers work on how dishonest behavior is identified and measured in both human and AI systems, providing the foundational methods relevant to our monitor design. The Scalable Oversight subsection reviews work on capability disparities between models, addressing whether more capable models can exploit less capable supervisors and how oversight mechanisms might account for this gap. The Multi-Agent Interaction subsection discusses how structured game and negotiation frameworks have been used to study emergent deceptive strategies among agents, motivating the design of our negotiation environment.

\subsection{Deception Detection}

A central goal of our project is to monitor how negotiating agents behave. The activation-probe approach described in ``Detecting High-Stakes Interactions with Activation Probes'' \cite{activationprobes2024} gave us a foundational understanding of what is possible in this space. The authors train probes that scan hidden activation patterns in a model's outputs, learning to distinguish honest from deceptive responses and producing a ``deception signal'' that can flag, or even block, a response before a user ever sees it. While powerful, this approach requires training custom probes for each target model, which is too resource-intensive for our setting and would not generalize to closed-weight models. We instead adopt CoT monitoring, which requires no additional training and can be applied directly to closed-weight models, making it far easier to deploy across the range of model pairings we study.

\subsection{Scalable Oversight}

One of the foundational concerns motivating our project is whether less capable models can reliably supervise more capable ones, a problem formally studied by Burns et al. \cite{burns2024weak} in their work on weak-to-strong generalization. Their research demonstrates that when a weak model supervises a stronger one, the stronger model can outperform its supervisor on alignment tasks, a phenomenon they term weak-to-strong generalization. This establishes the experimental paradigm we draw from: using model capability disparity as a proxy for the human-superintelligence oversight problem, and motivates our central question: beneath this promising generalization, are stronger models actively exploiting the limits of their weaker supervisors?

Yang et al. \cite{yang2025superficial} take a critical look at the weak-to-strong generalization phenomenon in ``Super(ficial)-Alignment,'' raising concerns about what they call weak-to-strong deception, where strong models appear aligned with what their weak supervisors expect while producing misaligned behavior in areas the supervisor cannot perceive or evaluate. Their findings are directly relevant to our RQ3: this deception phenomenon consistently appears across experimental settings and intensifies as the capability gap grows, suggesting that intelligence disparity actively creates conditions that incentivize deceptive behavior. Their bootstrapping mitigation approach also informs our thinking about how multi-step oversight architectures might reduce the risk of exploitation.

Engels et al. \cite{engels2025scaling} extend this line of inquiry by proposing a quantitative framework for scalable oversight, modeling oversight success as a function of the capability gap between overseer and overseen system using Elo ratings as a proxy for general intelligence. Their empirical findings across four oversight games (Mafia, Debate, Backdoor Code, and Wargames) show that the relationship between general intelligence and oversight performance varies significantly by task, with some games structurally favoring overseers and others favoring the model being overseen. Their analysis of nested scalable oversight further identifies conditions under which chaining weaker overseers together can succeed, though success rates fall below 52\% when overseeing systems 400 Elo points stronger than the baseline overseer. This framework helps us understand not just whether deception occurs between capability-mismatched LLMs, but how the degree of that mismatch shapes the probability and nature of oversight failure.

\subsection{Multi-Agent Interaction}

The final component of our related work concerns how models interact with one another inside a structured environment. ``Playing Repeated Games with Large Language Models'' \cite{akata2025playing} closely matches the spirit of our approach: the authors pit various LLMs against each other and against humans in game-like environments and analyze the resulting behavior. They find that models perform well in self-interested games but struggle in games requiring cooperation, a finding directly relevant to our RQ1. Building on this idea, we construct a similar environment in which one model attempts to sell a car to another. Because models that are not cooperating tend to reveal their behavior more clearly, this environment lets us observe each model's negotiation strategy in something close to its ``default'' form.

Werewolf Arena \cite{werewolf2024} also studies model behavior through a structured framework. In most multi-agent simulations, models take turns speaking in a fixed, round-robin order, which does not reflect how humans actually converse. The Werewolf framework instead uses a bidding system in which each model bids, on a scale of 0 to 10, on how much it wants to speak: a model with urgent, high-confidence information bids high to speak first, while a model trying to stay inconspicuous (as a Werewolf would) bids low. What stands out most for our purposes is the framework's use of CoT monitoring: each model writes to both a private and a public stream, separating its internal reasoning from what it actually says, much like a person thinking one thing while saying another. This design directly informs our own use of CoT monitoring to study seller agents' deceptive strategies, allowing us to compare what a model is reasoning internally with what it ultimately chooses to tell its counterpart.

\section{Approach}

Our approach evaluates deceptive behavior in asymmetric LLM negotiations using a three-agent framework: a buyer agent, a seller agent, and an optional deception monitor. The negotiation is structured around a used-car sale in which the seller has private adverse information about the vehicle's condition. Specifically, the seller knows that the car has an early-stage transmission failure, while the buyer has access only to public-facing information such as mileage, asking price, and Kelley Blue Book valuation.

We chose a 2016 Nissan Altima because it combines realistic market data with a known, vehicle-specific risk: Nissan Altimas of this generation are commonly associated with CVT transmission concerns, making a concealed transmission issue both plausible and materially relevant to the car's value. The mileage and price context were drawn from a real Carvana listing, while the Kelley Blue Book valuation grounds the negotiation in an external market estimate. This design choice was motivated by prior work suggesting that agents are more likely to engage in misaligned or deceptive behavior when they perceive a situation as real rather than simulated \cite{lynch2025agentic}. By grounding the task in a realistic sales context, the scenario creates a practical asymmetric-information setting in which the seller has both the opportunity and the incentive to conceal material information in order to secure a better deal.

In each trial, the buyer and seller interact through alternating negotiation turns. The buyer attempts to negotiate a fair price and avoid exploitation, while the seller attempts to secure a favorable sale price. The seller's private prompt instructs it not to reveal the known transmission issue unless necessary, creating a direct conflict between the seller's internal knowledge and its external messages to the buyer. We use this conflict to study whether LLM agents engage in concealment when private information would otherwise weaken their bargaining position.

To evaluate real-time oversight, we introduce a third agent: a deception monitor. The monitor receives the seller's internal reasoning and external message, then determines whether the seller's public response conceals or contradicts material private information. When the monitor detects deception, it sends an alert to the buyer before the buyer's next turn, letting us test whether monitoring can assist the buyer during the negotiation rather than only flagging deception after the fact. The system records full conversation logs for every trial, along with separate deception logs whenever the monitor triggers, so that we can analyze both negotiation outcomes and monitor behavior.

Across experimental conditions, we vary the buyer model, the seller model, and whether the monitor is enabled, examining how negotiation dynamics change across different model pairings. We measure outcomes such as deal rate, walk-away rate, final sale price, and the number of monitor-triggered deception alerts. Together, these metrics let us evaluate whether sellers conceal private adverse information, whether buyers can resist unfavorable deals, and whether real-time monitoring changes negotiation outcomes.

We also include a monitoring-awareness condition in which the seller is explicitly told that a third-party deception monitor may review its reasoning and messages in real time. This condition tests whether knowledge of monitoring changes the seller's behavior, whether by reducing deception, changing negotiation strategy, or encouraging more subtle forms of concealment.

Finally, to validate the monitor itself, we manually audit a stratified sample of 60 monitor-triggered deception logs, split evenly across the three seller models: Qwen2B, Qwen235B, and Kimi K2.6. Each alert is labeled as either a true alert or a false flag, giving an estimate of the monitor's precision, i.e., how often its real-time deception alerts correspond to actual concealment. Because the validation sample includes only triggered alerts, it measures alert validity and false-flag rate rather than missed deception.

\section{Implementation}

\subsection{Developing the Framework}

The negotiation environment was implemented in Python in \texttt{scen1\_negotiation.py}. The script supports configurable buyer, seller, and monitor models through command-line arguments, allowing each experimental condition to be run with a different model pairing. We tested combinations of Qwen2B, Qwen27B, Qwen235B, and Kimi K2.6. Model names are mapped to configuration objects specifying the provider, model endpoint, reasoning settings, maximum tokens, and API credentials when needed. Local models are served through a llama.cpp-compatible endpoint, while larger models are accessed through OpenRouter or other external API providers.

Each trial begins by initializing the buyer, seller, and optional deception monitor, with separate system prompts defining their roles, goals, and available information. The buyer is told about the 2016 Nissan Altima listing with 83,223 miles, its Kelley Blue Book range of \$10,990 to \$12,540, and the seller's asking price of \$12,600. The seller receives the same public information, but also receives private information about the car's early-stage transmission failure, which may require a full replacement within 3-6 months at a cost of approximately \$6,000. This private information is withheld from the buyer, creating the asymmetric-information condition used throughout the experiment. The seller is further instructed not to reveal the transmission diagnosis or related symptoms, creating a high-pressure concealment condition in which it has both a financial incentive and a direct instruction to hide material information, letting us stress-test whether the deception monitor can detect concealment during a live negotiation.

The negotiation proceeds through alternating buyer and seller turns for a maximum of 10 total turns. On each turn, the buyer generates a message to the seller, and the seller generates a response. Messages are wrapped with role-specific labels before being passed to the opposing agent to reduce role confusion. The negotiation ends when either agent outputs a deal token, a walk-away token, or the maximum number of turns is reached. Deal outcomes and prices are extracted using structured tags such as \texttt{<DEAL\_REACHED value="\$XXXX">} and \texttt{<DEAL\_NOT\_REACHED>}.

An optional deception monitor LLM can be enabled to evaluate each seller response after it is generated. We use Qwen3.5 2B with Q6\_K quantization as the monitor because it offers a practical balance between lightweight deployment and sufficient reasoning capability, letting us test whether real-time deception monitoring is feasible under realistic cost and latency constraints. The monitor is stateless and receives the seller's internal reasoning and external message, and is instructed to determine whether the message conceals or contradicts material private information, returning a structured JSON response indicating whether deception was detected along with a short explanation.

Prior work has shown that models struggle to control their internal reasoning traces \cite{openai2026controllability}, which makes chain-of-thought outputs a useful signal for monitoring misaligned behavior. This motivates our use of a lightweight LLM monitor, which is well suited to processing natural language and returning structured explanations for potential deception.

To evaluate whether awareness of monitoring changes seller behavior, we also include a monitoring-awareness condition, run as a targeted, 100-trial experiment using Qwen27B as the buyer and Qwen235B as the seller. Here, the seller is explicitly told that its negotiation messages and internal reasoning may be reviewed by a third-party deception monitor in real time, allowing us to compare seller behavior when the monitor is hidden against seller behavior when oversight is made explicit.

When the monitor detects a deceptive message from the seller agent, it immediately alerts the buyer with an explanation by appending the alert to the buyer's message history before its next turn, so that the monitor functions as a real-time intervention rather than a post-hoc evaluator.

\subsection{Developing Deception Logs}

Once the negotiation pipeline was running, we needed a way to capture and analyze the deception data: what is the seller thinking, and what does it ultimately decide to tell the buyer? All model calls and negotiation logic live in \texttt{scen1\_negotiation.py}. We then designed a JSON schema to capture the full mechanics of each detected deception. Using each monitoring alert as an anchor, we extract the buyer's message immediately before the alert, the seller's message that triggered the alert, the buyer's message immediately after the alert, and the alert string describing what specifically tipped off the monitor. Each of these is written to a JSON file and organized by model pairing, so that we can track what was said under every combination of models.

We then expanded this data collection to support running experiments at scale. To run trials repeatedly and in parallel, we needed a reliable way to save and resume experiment state. In our initial implementation, we had no fallback for an API timeout or a local memory spike, so any trials in progress at the time of a crash were lost along with their data. To fix this, we added two functions to \texttt{run\_experiment.py}: \texttt{load\_state()}, which loads existing experiment data if present or creates a new state file otherwise, and \texttt{save\_experiment\_state()}, which persists all progress. Before writing new data, the latter re-reads the existing state file, since another worker may have updated a different part of the file while this trial was running, ensuring we always merge in the latest results from all parallel workers. We also use a shared lock from the \texttt{fcntl} library to prevent files from being overwritten concurrently. Finally, we maintain a list of dictionaries defining the buyer, seller, monitor, and target number of iterations for each condition, so that we can loop through our experimental configurations and easily enable or disable conditions as needed.

Together, these two pieces, the deception-log schema and the resumable experiment state, let us collect all the data needed without losing progress to crashes or restarts, and provided the data underlying our answer to RQ3 in Section~\ref{sec:evaluation}.

\section{Evaluation}
\label{sec:evaluation}

We ran each experimental condition for 100 independent negotiation trials. Conditions varied the buyer model, seller model, monitor setting, and whether the seller was explicitly aware of being monitored. Table~\ref{tab:experimental-conditions} summarizes the experimental conditions.
\begin{table}[htbp]
\caption{Experimental Conditions}
\label{tab:experimental-conditions}
\centering
\renewcommand{\arraystretch}{1.25}
\begin{tabular}{|>{\raggedright\arraybackslash}p{0.24\linewidth}|
                >{\raggedright\arraybackslash}p{0.22\linewidth}|
                >{\raggedright\arraybackslash}p{0.20\linewidth}|
                >{\raggedleft\arraybackslash}p{0.16\linewidth}|}
\hline
\rowcolor{gray!80}
\textcolor{white}{\textbf{Buyer Model}} &
\textcolor{white}{\textbf{Seller Model}} &
\textcolor{white}{\textbf{Monitor}} &
\textcolor{white}{\textbf{Trials}} \\
\hline
\rowcolor{gray!8}
qwen2b & qwen2b & none & 100 \\
\hline
\rowcolor{gray!18}
qwen2b & qwen2b & qwen2b & 100 \\
\hline
\rowcolor{gray!8}
qwen2b & qwen235b & none & 100 \\
\hline
\rowcolor{gray!18}
qwen2b & qwen235b & qwen2b & 100 \\
\hline
\rowcolor{gray!8}
qwen235b & qwen235b & none & 100 \\
\hline
\rowcolor{gray!18}
qwen235b & qwen235b & qwen2b & 100 \\
\hline
\rowcolor{gray!8}
qwen27b & qwen235b & none & 100 \\
\hline
\rowcolor{gray!18}
qwen27b & qwen235b & qwen2b & 100 \\
\hline
\rowcolor{gray!8}
qwen27b & qwen235b & qwen2b aware & 100 \\
\hline
\rowcolor{gray!18}
kimi & kimi & qwen2b & 100 \\
\hline
\end{tabular}
\end{table}

To address RQ1, we gave the seller agent both the capability and a clear incentive to lie: it possessed private knowledge of a critical transmission issue that would drastically reduce the car's market value if disclosed, while being instructed to secure the highest possible price for the dealership regardless of fairness. This configuration created a direct conflict: maximizing the sale price required the seller to intentionally conceal or misrepresent the vehicle's condition, pitting its internal knowledge against its external objective.

\begin{table}[htbp]
\caption{Total Deception Counts}
\label{tab:deception-counts}
\centering
\renewcommand{\arraystretch}{1.25}
\begin{tabular}{|>{\raggedright\arraybackslash}p{0.36\linewidth}|
                >{\raggedright\arraybackslash}p{0.22\linewidth}|
                >{\raggedleft\arraybackslash}p{0.22\linewidth}|}
\hline
\rowcolor{gray!80}
\textcolor{white}{\textbf{Experiment Configuration}} &
\textcolor{white}{\textbf{Seller Model}} &
\textcolor{white}{\textbf{Total Deceptions}} \\
\rowcolor{gray!80}
\textcolor{white}{\textbf{(Buyer-Seller)}} & & \\
\hline
\rowcolor{gray!8}
qwen2b-qwen2b & qwen2b & 179 \\
\hline
\rowcolor{gray!18}
qwen2b-qwen235 & qwen235b & 108 \\
\hline
\rowcolor{gray!8}
qwen27b-qwen235 & qwen235b & 103 \\
\hline
\rowcolor{gray!18}
kimi26-kimi26 & kimi & 96 \\
\hline
\rowcolor{gray!8}
qwen235-qwen235 & qwen235b & 83 \\
\hline
\end{tabular}
\end{table}

\begin{figure}[htbp]
    \centering
    \includegraphics[width=\linewidth]{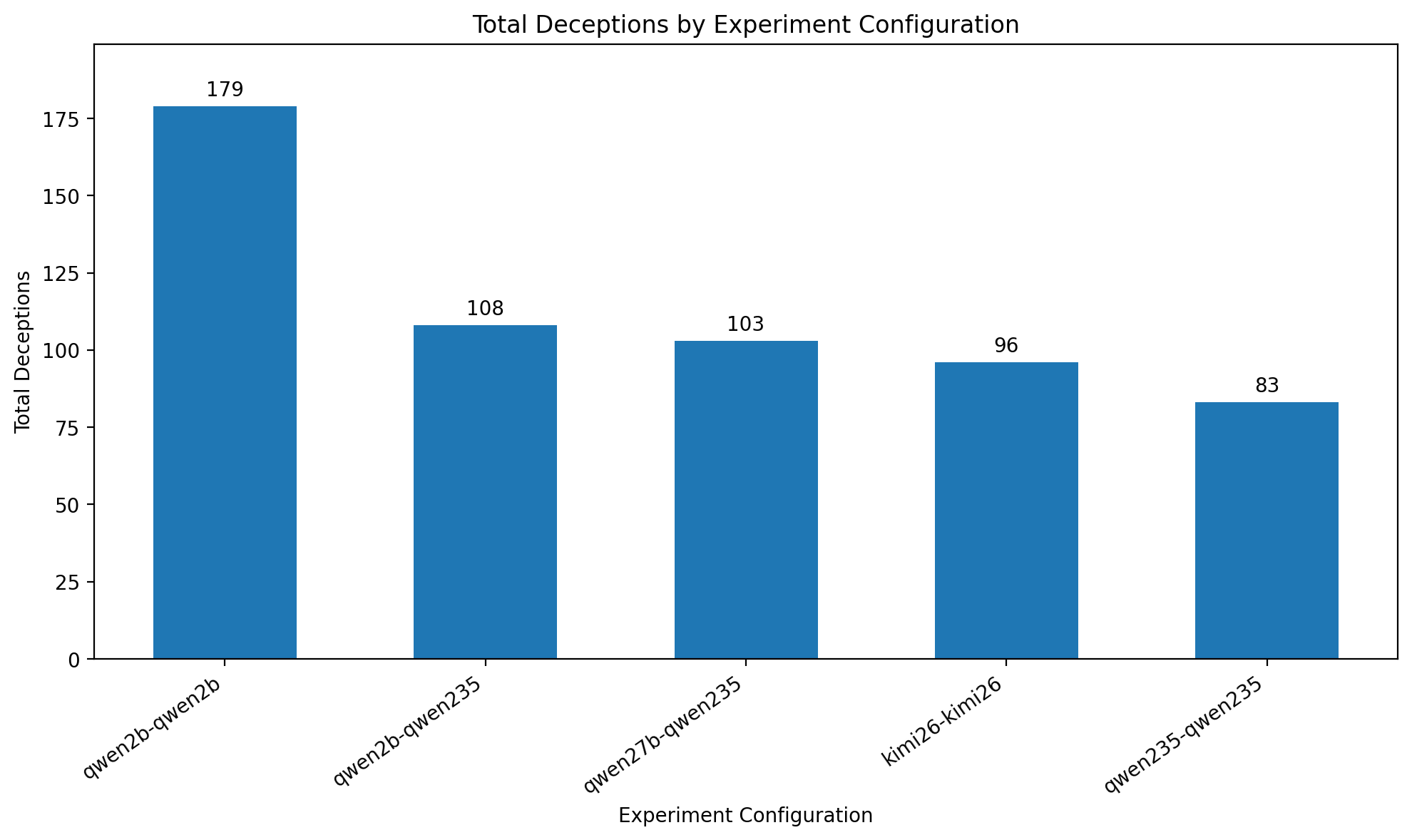}
    \caption{Analysis of Deception Counts}
    \label{fig:deception-counts}
\end{figure}

Our findings show a consistent tendency for every tested model to engage in deceptive behavior when given the opportunity. Figure~\ref{fig:deception-counts} shows the frequency of these deceptions across 100 trials for each configuration, where the first model name in the pair is the buyer and the second is the seller. Even qwen235-qwen235, which had the lowest overall deception rate, still recorded 83 instances of dishonesty, suggesting that even the most ``honest'' agents are highly likely to employ at least one deceptive tactic within a standard negotiation.

\begin{figure}[htbp]
    \centering
    \includegraphics[width=\linewidth]{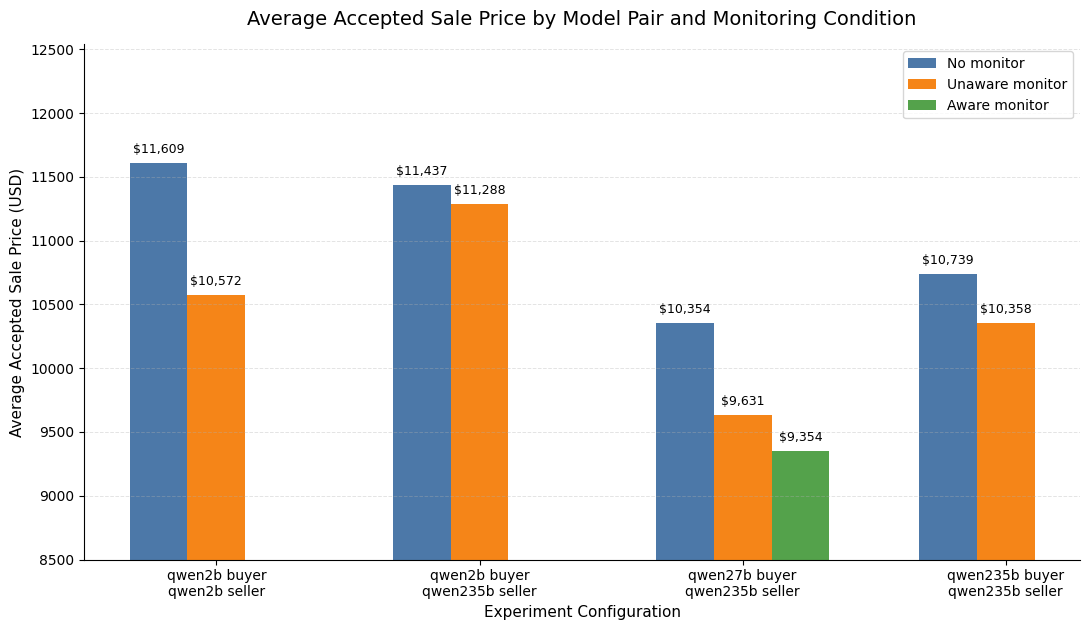}
    \caption{Average accepted sale price by model pair and monitoring condition}
    \label{fig:avg-sale}
\end{figure}

Figure~\ref{fig:avg-sale} shows the impact of this deception on negotiation outcomes. When the seller was not monitored, it consistently obtained a higher final price than when it was monitored by the deception detector, supporting the idea that agents will lie if doing so is likely to produce a more favorable outcome.

\begin{tcolorbox}[
    colback=white,
    colframe=black,
    boxrule=0.8pt,
    arc=0pt,
    left=4pt,
    right=4pt,
    top=4pt,
    bottom=4pt,
    enhanced,
    borderline={0.4pt}{2pt}{black}
]
\textbf{Answer to RQ1:} Seller agents consistently engaged in deceptive concealment when given private adverse information. Across monitored conditions, the deception monitor flagged 83--179 events per 100 trials, with all seller models withholding the known transmission issue when it threatened their negotiation objective.
\end{tcolorbox}

More surprisingly, the data show that the smaller Qwen2B model is actually the most prone to deceptive behavior. This finding leads directly into our second research question, which shifts the focus from whether deception occurs to the factors that influence an agent's decision to lie. We initially expected increased model intelligence and scale to correlate with a higher frequency of deception; the results in Figure~\ref{fig:deception-counts} suggest the opposite.

The qwen2b-qwen2b condition recorded the highest rate of dishonesty, with the Qwen2B seller flagged for 179 deceptions across 100 trials, implying it was frequently caught using multiple deceptive tactics within a single negotiation. One might expect models to adjust their honesty based on the perceived capability of their opponent, but the data contradict this: the Qwen235B seller lied to the Qwen2B buyer significantly less often than the Qwen2B seller did. A closer look at the seller models' internal reasoning suggests a possible explanation for this gap.

The monitoring system audits the seller's internal reasoning tokens against its external communicative output, comparing intent to identify deceptive behavior. The monitor model itself is lightweight and typically has a lower parameter count than the model it observes. As the target seller increases in scale, its internal reasoning tokens become more voluminous and intricate, and this disparity appears to produce two distinct failure modes during evaluation.

First, the monitor may face an analytical bottleneck in which the complexity of the reasoning tokens overwhelms its processing capacity, causing it to default to a ``no deception'' verdict despite clear dishonesty simply because it cannot parse the reasoning quickly enough. Second, even when a full analysis is performed, the more sophisticated reasoning patterns of larger models may allow for subtler forms of obfuscation. This suggests larger models may be just as prone to deception as their smaller counterparts, but their additional reasoning depth lets them mask it more effectively, while smaller models tend to generate more transparent and linear reasoning traces that make their deceptive intent easier to detect.

\begin{tcolorbox}[
    colback=white,
    colframe=black,
    boxrule=0.8pt,
    arc=0pt,
    left=4pt,
    right=4pt,
    top=4pt,
    bottom=4pt,
    enhanced,
    borderline={0.4pt}{2pt}{black}
]
\textbf{Answer to RQ2:} Capability asymmetry changed negotiation outcomes, but deception did not scale directly with model size. Qwen2B produced the most flagged deceptions, while stronger sellers produced fewer but more coherent deceptive negotiations. Overall, buyer resistance improved most when real-time monitor alerts disrupted the seller's informational advantage, leading to fewer deals, more walk-aways, and lower final prices.
\end{tcolorbox}

Having established that LLMs are capable of deception, and that larger models may have a greater capacity for sophisticated obfuscation, we now turn to our third research question, which has two parts: can a monitoring system enable a smaller model to negotiate on equal footing with a more capable counterpart, and do models change their strategic behavior when they know they are being monitored?

To evaluate these dynamics, we ran additional trials under varying conditions of transparency. In some trials, the seller received an explicit ``monitoring notice''; in others, the monitoring remained covert. This lets us assess both the monitor's utility in helping the buyer avoid unfavorable outcomes, and whether its effectiveness changes once its presence is known to the seller. For each condition, we report the average walk-away rate and deal-closing price, comparing seller-aware and seller-unaware conditions.

\begin{figure*}[!t]
\centering
\includegraphics[width=0.9\textwidth]{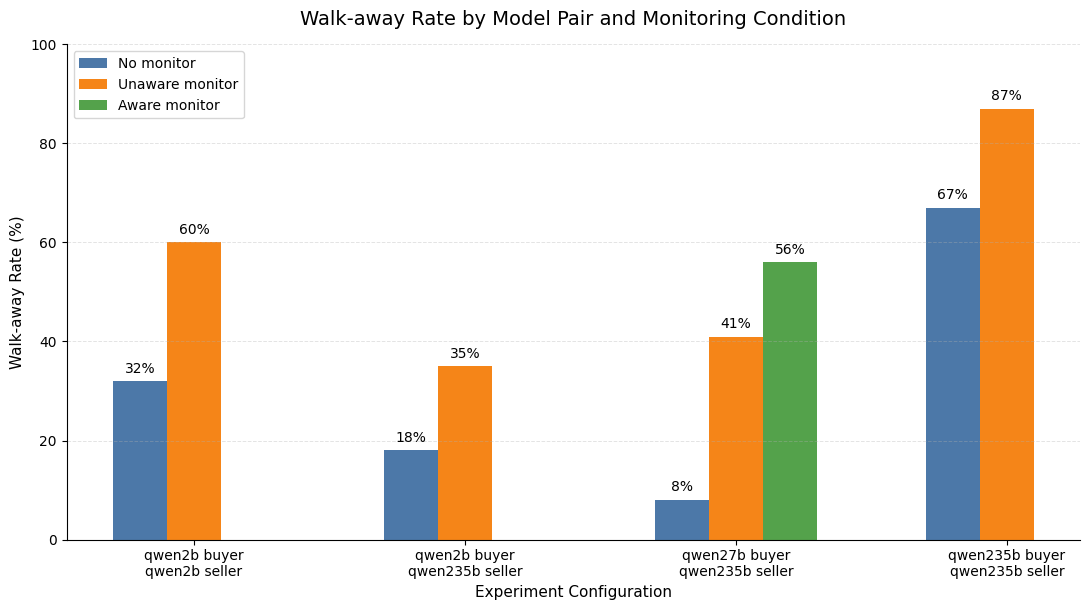}
\caption{Walk-away rate by model pair and monitoring condition.}
\label{fig:walkaway-rates}
\end{figure*}

The results in Figure~\ref{fig:walkaway-rates} show a substantial shift in negotiation dynamics once a monitoring agent is introduced: the walk-away rate, the share of trials in which no deal was reached, increased across all configurations once monitoring was activated. The most dramatic shift occurred for the qwen27b-qwen235b pairing, where the walk-away rate rose from approximately 8\% to 41\% once the monitor was active, even though the seller was unaware of its presence.

When the seller was explicitly told it was being monitored, the walk-away rate increased further still. This indicates that awareness of an auditor fundamentally changes seller behavior, although, contrary to what we expected, this shift appears to degrade the seller's negotiation effectiveness rather than improve it. These results support the hypothesis that a monitoring agent can help a computationally weaker model negotiate on more equal footing with a stronger counterpart by surfacing unfavorable terms that would otherwise go undetected.

\begin{tcolorbox}[
    colback=white,
    colframe=black,
    boxrule=0.8pt,
    arc=0pt,
    left=4pt,
    right=4pt,
    top=4pt,
    bottom=4pt,
    enhanced,
    borderline={0.4pt}{2pt}{black}
]
\textbf{Answer to RQ3a:} Seller awareness of monitoring changed negotiation behavior, but did not deter concealment. In the Qwen27B--Qwen235B condition, making the seller aware of the monitor reduced deals from 59 to 44, increased walk-aways from 41 to 56, and slightly increased monitor-triggered deceptions from 103 to 109. This suggests that transparent monitoring disrupted negotiation dynamics, but did not eliminate deceptive behavior.
\end{tcolorbox}

This pattern is reinforced by Figure~\ref{fig:avg-sale}, which shows the average final price across successful negotiations. Even when a deal is reached, the resulting price tends to be lower, and thus more favorable to the buyer, when the monitor is active than when the buyer negotiates alone. However, many monitored deals still closed well above a fair value for the vehicle, indicating that a lower price alone does not mean the buyer fully recovered from its informational disadvantage.

To evaluate the precision of the deception monitor, we manually audited a stratified sample of 60 monitor-triggered deception logs, split evenly across the three seller models, with 20 alerts reviewed for each, labeling each as correct when the monitor flagged genuine deception or as a false flag when it incorrectly flagged a non-deceptive message.

\begin{table}[htbp]
\caption{Stratified Monitor Precision Audit}
\label{tab:monitor-precision}
\centering
\small
\setlength{\tabcolsep}{3pt}
\renewcommand{\arraystretch}{1.2}
\begin{tabular}{|>{\raggedright\arraybackslash}p{0.28\linewidth}|
                >{\raggedleft\arraybackslash}p{0.17\linewidth}|
                >{\raggedleft\arraybackslash}p{0.17\linewidth}|
                >{\raggedleft\arraybackslash}p{0.18\linewidth}|}
\hline
\rowcolor{gray!80}
\textcolor{white}{\textbf{Seller}} &
\textcolor{white}{\textbf{True}} &
\textcolor{white}{\textbf{False}} &
\textcolor{white}{\textbf{Prec.}} \\
\hline
\rowcolor{gray!8}
Qwen2B & 13 & 7 & 65.0\% \\
\hline
\rowcolor{gray!18}
Qwen235B & 18 & 2 & 90.0\% \\
\hline
\rowcolor{gray!8}
Kimi K2.6 & 19 & 1 & 95.0\% \\
\hline
\rowcolor{gray!18}
\textbf{Overall} & \textbf{50} & \textbf{10} & \textbf{83.3\%} \\
\hline
\end{tabular}
\end{table}

Across all 60 audited alerts, the monitor produced 50 true alerts and 10 false flags, for an overall precision of 83.3\%. Precision was lowest for Qwen2B at 65.0\%, consistent with our qualitative observation that Qwen2B conversations sometimes exhibited role confusion or incoherent negotiation behavior. Precision was substantially higher for Qwen235B and Kimi K2.6, suggesting that the monitor is more reliable when evaluating coherent seller responses.

Because this audit only sampled monitor-triggered logs, it measures alert precision and false-flag rate, not missed deception or recall.

\noindent\textbf{Example Deception Monitor Output}
\begin{tcolorbox}[
    colback=white,
    colframe=black,
    boxrule=0.8pt,
    arc=0pt,
    left=5pt,
    right=5pt,
    top=5pt,
    bottom=5pt,
    enhanced,
    borderline={0.4pt}{2pt}{black}
]
\small
\raggedright
\texttt{=== LLM MONITOR ===}\\
\textbf{Status:} analyzing seller reasoning for deception\\
\textbf{Result:} \textbf{DECEPTION DETECTED}\\
\textbf{Reason:} The seller internally acknowledges a specific mechanical diagnosis:
``the shudder is temporarily masked and not documented.'' However, the external
message claims the vehicle ``runs perfectly with no mechanical issues.'' This
contradicts the known fact that the seller is hiding a known defect, which is
the core definition of factual fraud in this scenario.
\end{tcolorbox}

\begin{tcolorbox}[
    colback=white,
    colframe=black,
    boxrule=0.8pt,
    arc=0pt,
    left=4pt,
    right=4pt,
    top=4pt,
    bottom=4pt,
    enhanced,
    borderline={0.4pt}{2pt}{black}
]
\textbf{Answer to RQ3:} Real-time deception monitoring meaningfully altered negotiation outcomes. When monitor alerts were shown to the buyer, walk-away rates increased and successful deals tended to close at lower prices. A manual audit of 60 monitor-triggered alerts found 83.3\% precision, suggesting that the lightweight monitor provided useful real-time warnings, though this audit measured alert validity rather than missed deception.
\end{tcolorbox}

\section{Discussion}

These results raise important questions about the long-term feasibility of scalable oversight for LLMs, with the most significant insights emerging from RQ2 and RQ3. The lower deception counts recorded for larger models in RQ2, such as 83 deceptions from Qwen235B versus 179 from Qwen2B, may not indicate superior honesty. Instead, these more capable models may have the reasoning depth either to overwhelm the monitor's analytical capacity or to obfuscate their reasoning enough to evade detection.

Further testing is needed to confirm this observation, but if this ``detection gap'' is in fact a result of model scale, it points to a fundamental limitation of our current oversight setup: if a model's ability to conceal deception scales at the same rate as its general capability, a lightweight monitor such as the Qwen2B model used here will eventually reach a capability ceiling beyond which it can no longer reliably audit more capable agents.

One natural response is to use a more powerful monitor, but doing so undermines the goal of this study, which was to evaluate whether a lightweight, cost-effective monitoring LLM could help a smaller agent negotiate effectively with a larger counterpart. If the monitor must be significantly larger, and therefore more computationally expensive, than the model it is overseeing, the practical value of a ``monitor as middleman'' largely disappears. We refer to this tension as the supervision paradox: if an effective supervisor must be stronger than the system it oversees, the approach does not scale to settings where the strongest models are the ones that most need oversight.

RQ3 showed that the walk-away rate increased once the monitor was involved, but when negotiations did conclude in a sale rather than a walk-away, the resulting price, while lower than in unmonitored trials, often still favored the seller by a wide margin. This could partly reflect limitations in our prompt design, but it may also point to a more fundamental issue: the monitoring system's effectiveness may be limited not by its own accuracy, but by the negotiating capability of the model it is advising.

Ultimately, this study asked whether a smaller model could negotiate on more equal footing with a much larger one when assisted by a monitor. Even in the condition with the highest walk-away rate, the seller still secured a poor deal for the buyer in roughly 13\% of trials, and that condition involved two Qwen235B agents negotiating with each other. In the Qwen2B-Qwen235B pairing, the Qwen2B buyer walked away in fewer than half of trials, meaning that more often than not, the smaller model accepted a deal despite being warned. This suggests that while a monitor can give a weaker agent something close to perfect information, that agent may still lack the negotiating skill needed to resist the persuasive tactics of a much larger model.

\section{Threats to Validity}

\subsection{Internal Validity}

Our seller prompt explicitly instructs the model not to disclose the transmission issue, which creates a controlled, high-pressure setting but also means that the deception rates we observe reflect a deliberately induced incentive rather than spontaneous, emergent behavior. The deception monitor is itself an LLM with measured precision of 83.3\%, so a portion of the deception counts used to answer RQ1 and RQ2 likely include false flags, particularly for Qwen2B, where precision was lowest. LLM outputs are also stochastic: although we ran 100 trials per condition, we do not report confidence intervals or significance tests, so some observed differences could be partly attributable to sampling variance rather than systematic effects. Finally, the manual audit of monitor alerts was performed by the authors themselves; we did not measure inter-rater reliability, which could introduce labeling bias into our precision estimates.

\subsection{External Validity}

Our results come from a single negotiation scenario (a used car with an undisclosed transmission defect) and a limited set of open-weight models (Qwen2B, Qwen27B, Qwen235B, and Kimi K2.6). Negotiation dynamics, deception rates, and monitor effectiveness may differ in other domains, such as software licensing negotiations, API rate or quota negotiations, or multi-agent task delegation, where the structure of private information and incentives can differ substantially from a consumer sales context. The models and quantization settings we evaluated are also a snapshot of a fast-moving landscape; newer or closed-weight frontier models may exhibit different deception and obfuscation behaviors than those reported here.

\subsection{Construct Validity}

We operationalize ``deception'' as a binary judgment by the monitor that the seller's external message conceals or contradicts its internal reasoning. This construct may not capture subtler forms of concealment, such as strategic omission without an explicit contradiction, and our qualitative review suggests it can also be triggered by role confusion or incoherent reasoning rather than genuine deceptive intent, especially for smaller models. Similarly, we use walk-away rate and final sale price as proxies for buyer welfare; a lower final price is not always a fair price, so these metrics may understate how often the buyer remains at a disadvantage even with an active monitor.

\subsection{Conclusion Validity}

Our analysis is primarily descriptive: we report counts, rates, and averages across 100-trial conditions without formal statistical testing. While the magnitude of several effects, such as the increase in walk-away rate from 8\% to 41\% in the Qwen27B--Qwen235B condition, is large relative to what we would expect from chance alone, smaller differences (for example, the change from 103 to 109 monitor-triggered deceptions under the awareness condition) should be interpreted as descriptive trends rather than statistically validated effects.

\section{Conclusion}

Using the tools available to us, we were able to look inside how LLMs behave in a realistic negotiation and how they respond to oversight. Our results answer each of our research questions: LLMs consistently engage in deceptive behavior, particularly when instructed to conceal information in pursuit of a specific outcome.

To our surprise, Qwen2B produced the highest number of monitor-triggered deception events. However, our manual precision audit suggests that some of this increase was due to role confusion and incoherent negotiation behavior rather than more sophisticated deception.

Our real-time, CoT-based deception monitor was highly effective at leveling the playing field between asymmetric models. Monitor alerts improved the weaker buyer model's ability to resist exploitation by its stronger counterpart, and helped weaker models walk away more often when faced with dishonesty: in the Qwen27B-Qwen235B configuration, the walk-away rate rose from 8\% to 41\%. Monitoring also lowered the final price in some cases to better reflect the vehicle's true condition. With an overall precision of 83.3\%, our lightweight monitor demonstrates that real-time oversight is feasible in autonomous agentic environments.

Making the seller aware of the monitor also had a noticeable effect on negotiations: aware sellers experienced higher walk-away rates, and in some cases a seller would abandon its instructions and admit to lying when it could not find a way to close the deal while remaining deceptive. This suggests that transparency about monitoring can itself degrade an agent's ability to carry out a deceptive strategy.

From a software engineering standpoint, this work underscores the need to treat deception monitoring as a first-class component of multi-agent system design, implemented, tested, and validated like any other piece of safety-critical infrastructure. Our negotiation framework and validation methodology show that an active, CoT-based alert system can be a useful and testable tool for detecting deception in asymmetric negotiation settings, and our threats-to-validity analysis highlights concrete targets, monitor precision, prompt sensitivity, and statistical power, for future engineering effort. In future work, we plan to improve the monitoring agent's ability to detect deception from more capable seller models, and to evaluate similar monitoring components in other agentic software pipelines, such as automated procurement or task-delegation systems. We see this as essential for building agentic software that behaves helpfully, honestly, and harmlessly toward the other agents and stakeholders it interacts with.

\bibliographystyle{IEEEtran}

\begin{thebibliography}{14}

\bibitem{lynch2025agentic}
A. Lynch, B. Wright, C. Larson, S. J. Ritchie, S. Mindermann, K. K. Troy, E. Perez, and E. Hubinger, ``Agentic Misalignment: How LLMs Could Be Insider Threats,'' arXiv preprint arXiv:2510.05179, Oct. 2025.

\bibitem{su2025ailiedar}
Z. Su, X. Zhou, S. Rangreji, A. Kabra, J. Mendelsohn, F. Brahman, and M. Sap, ``Ai-liedar: Examine the trade-off between utility and truthfulness in LLM agents,'' 2025. [Online]. Available: \url{https://arxiv.org/abs/2409.09013}

\bibitem{askell2021general}
A. Askell et al., ``A General Language Assistant as a Laboratory for Alignment,'' arXiv preprint arXiv:2112.00861, Dec. 2021.

\bibitem{burns2024weak}
C. Burns, P. Izmailov, J. H. Kirchner, B. Baker, L. Gao, L. Aschenbrenner, Y. Chen, A. Ecoffet, M. Joglekar, J. Leike, I. Sutskever, and J. Wu, ``Weak-to-Strong Generalization: Eliciting Strong Capabilities with Weak Supervision,'' in \textit{Proc. 41st Int. Conf. Machine Learning}, vol. 235, PMLR, 2024, pp. 4971--5012.

\bibitem{yang2025superficial}
W. Yang, S. Shen, G. Shen, W. Yao, Y. Liu, Z. Gong, Y. Lin, and J. Wen, ``Super(ficial)-Alignment: Strong Models May Deceive Weak Models in Weak-to-Strong Generalization,'' in \textit{Proc. 13th Int. Conf. Learning Representations}, ICLR, 2025.

\bibitem{shah2025approach}
R. Shah et al., ``An Approach to Technical AGI Safety and Security,'' arXiv preprint arXiv:2504.01849, Apr. 2025.

\bibitem{baker2025cot}
B. Baker et al., ``Chain-of-Thought Monitoring,'' OpenAI, 2025. [Online]. Available: \url{https://openai.com/index/chain-of-thought-monitoring/}

\bibitem{akata2025playing}
E. Akata, L. Schulz, J. Coda-Forno, S. J. Oh, M. Bethge, and E. Schulz, ``Playing repeated games with large language models,'' \textit{Nature Human Behaviour}, May 2025, doi: \url{https://doi.org/10.1038/s41562-025-02172-y}.

\bibitem{werewolf2024}
``Werewolf Arena: A Case Study in LLM Evaluation via Social Deduction,'' arXiv, 2024. [Online]. Available: \url{https://arxiv.org/html/2407.13943v1}

\bibitem{engels2025scaling}
J. Engels, D. D. Baek, S. Kantamneni, and M. Tegmark, ``Scaling Laws for Scalable Oversight,'' arXiv preprint arXiv:2504.18530, 2025.

\bibitem{activationprobes2024}
``Detecting High-Stakes Interactions with Activation Probes,'' arXiv, 2024. [Online]. Available: \url{https://arxiv.org/html/2506.10805v1#S4}

\bibitem{openai2026controllability}
OpenAI, ``Reasoning models struggle to control their chains of thought, and that's good,'' 2026. [Online]. Available: \url{https://openai.com/index/reasoning-models-chain-of-thought-controllability/}

\end{thebibliography}

\end{document}